%Paper: hep-th/9310071
%From: devchand@thsun1.jinr.dubna.su (Chand Devchand)
%Date: Wed, 13 Oct 93 13:06:54 MSK

%Berezin.tex in plain tex
\magnification=1200
\parskip 3 pt plus 1pt minus 1 pt
\parindent=0pt
\overfullrule=0pt
\def\p{\partial} \def\D{{\cal D}}  \def\bD{{\bar {\cal D}}}
  
\def\op{\oplus} \def\om{\ominus}
\def\dpp{D^{++} }\def\dopp{D^{\op\op} }
\def\vpp{V^{++} }\def\vopp{V^{\op\op} }
\def\Dpp{\D^{++} }\def\Dopp{\D^{\op\op} }
\def\e{\epsilon}  \def\n{\nabla} \def\d{\delta}
    
\def\c{\chi} \def\t{\vartheta}  \def\bt{{\bar \vartheta}}
\def\b{\beta} \def\a{\alpha} \def\l{\lambda}  \def\f{\varphi}
\def\da{{\dot \alpha}} \def\db{{\dot \beta}}  
\def\sd{self-dual } \def\asd{anti-self-dual } \def\nsd{non-self-dual }
 \def\sdy{self-duality } \def\ym{Yang-Mills }
\def\sdym{self-dual Yang-Mills }
\def\ssdy{super self-duality } \def\sym{super Yang-Mills }
 \def\susy{supersymmetric }\def\eqs{equations }
\def\hs{harmonic superspace }
\def\half{{1\over 2}}
\def\der#1{{\partial \over \partial #1}}
\font\sqi=cmssq8
\def\R{\rm I\kern-1.45pt\rm R}
\def\C{\kern2pt {\hbox{\sqi I}}\kern-4.2pt\rm C}
\rightline{hep-th/9310071}
\rightline{Bonn-HE-93-32}
\rightline{Dubna-E2-93-361}
\vskip 1 true cm
\centerline{Integrability of N = 3 super Yang-Mills equations
\footnote{*}{to appear in the F. A. Berezin memorial volume}}
\vskip 1 true cm
\centerline{Ch. Devchand$^1$  and  V.  Ogievetsky$^{2,}
$\footnote{**}{ permanent address: Joint Institute for Nuclear
Research, Dubna, Russia}}
\vskip 15pt
\centerline{{\it  $^1$Joint Institute for Nuclear Research      }}
\centerline{{\it          Dubna, Russia            }}
\vskip 8pt
\centerline{{\it$^2$ Physikalisches Institut der Universit\"at Bonn }}
\centerline{{\it        Bonn, Germany }}
\vskip 2 true cm
\leftline{{\bf Abstract}}
We describe the harmonic superspace formulation of the Witten-Manin
supertwistor correspondence for N=3 extended super Yang-Mills theories.
The essence is that on being sufficiently supersymmetrised (up to the N=3
extension), the Yang-Mills equations of motion can be recast in the form of
Cauchy-Riemann-like holomorphicity conditions for a pair of prepotentials
in the appropriate harmonic superspace. This formulation makes the explicit
construction of solutions a rather more tractable proposition than previous
attempts.
\vskip 1 true cm
%\vfill \eject
\vskip 15pt{\bf 1. Introduction}\vskip 15pt
Alik Berezin was enthusiastic about the possibility of solving the
\ym equations and he frequently discussed this intriguing problem.
Recalling these discussions, we feel that he would have enjoyed knowing
about our recent work on this theme, which we shall describe  here as our
contribution to this volume dedicated to his memory.
Today the above possibility certainly shows much promise and many existence
proofs exist (e.g. [1]).
An older promise based on the twistor transform [2,3,4] has also
been renewed recently [5]; and this will be the subject of the present paper.
The latter approach was based on the observation that
one could approach \nsd \ym fields by combining \sd (SD) and \asd (ASD)
fields in some way. Specifically, the twistor transform for \sdym [6]
establishes a correspondence between \sd fields and certain holomorphic
fields, which effectively linearises the (nonlinear) \sdy \eqs, and the
observation of [2,3] concerned the intermingling of SD and ASD holomorphic
data to extend to \nsd data. In general, the procedure works in a certain
formal neighbourhood of the \sd solution, but Witten [2] observed that
for the N=3 \susy \eqs the SD and ASD data actually interlock to give an
exact \nsd solution. This observation
was given a global formulation by Manin [7], who also discussed some
solutions for rather complicated gauge groups.

Restricting ourselves entirely to local considerations, we shall show
that this supertwistorial construction enjoys a very elegant formulation
in the language of `harmonic superspace', which promises to be very
effective for the explicit construction of local solutions.
The crux of the formulation is that the $N=3$ extended supersymmetric
Yang-Mills equations can be rewritten as Cauchy-Riemann-like conditions
for a pair of prepotentials in an appropriate \hs. `Holomorphic'
prepotentials therefore encode local N=3 \sym solutions; leaving only the
decodification as the remaining technical problem.
The formulation involves a crucial modification of the harmonic (super)space
formulation of (anti-)\sdy equations (see, e.g. [8, 9] and previous
references references therein). The latter involve the harmonisation of
``half" the Lorentz group and only allow one to deal with base spaces of
signatures (4,0), (2,2) or with complexified space. For \nsd $N=3$, however,
we need to harmonise the {\it whole} Lorentz group, allowing us to consider
a base space of any signature, including (3,1), the Minkowski one .

We should mention that harmonics were originally introduced [10] in order
to construct the first unconstrained off-shell  $N = 2$ and $N = 3 $
\susy gauge theories. In that case the internal $SU(2)$ and $SU(3)$
groups were harmonised instead of the Lorentz group which is harmonised here.
Harmonisation of the three-dimensional Lorentz group  was discussed
for $N = 6$ $d = 3$ gauge theories by Zupnik [11].
\vskip 15pt{\bf 2. N=3 \sym \eqs }\vskip 15pt
The thrice-extended Yang-Mills multiplet contains the following
fields [12]: the gauge vector field represented by its self- and anti-dual
field-strengths $f_{\da \db}$ and $ f_{\a \b}$, defined by
$$ [\n_{\a\da} ~,~\n_{\b\db}]
\equiv\ \e_{\a\b} f_{\da\db}  ~+~ \e_{\da\db} f_{\a\b} ,  $$
spinor singlet and triplet fields $\{ \l_\a, \l_\da, \c^i_\da, \c_{i\a}\}$
and two triplets of scalar fields $W^i, W_i $ where $\a$ and $\da$ are
undotted and dotted Lorentz spinor indices while $i = 1, 2, 3$ is the
$SU(3)$ index. All fields are in the adjoint representation of gauge
group and are Lie-algebra-valued.
The dynamical equations for this supermultiplet are [13]
$$\eqalign{
\n^\a_{\ \db} f_{\b\a} +  \n_{\b}^{\ \da} f_{\da\db} &=\
 \{ \c^k_{\b} , \c_{k\db} \} + \{ \l_\b  , \l_\db \}
+ [ W^i , \n_{\b\db} W_i ]  +  [ W_i , \n_{\b\db} W^i ] \cr
\n^\b_{\ \db} \l_\b  &=\   [ \c_{i\db} , W^i ]  \cr
\n_{\a}^{\ \db} \l_\db  &=\  [ \c^i_{\a} , W_i ]  \cr
\n^\a_{\ \da} \c^j_\a  &=\    [ \c_{i\db} , W_k ] \e^{ijk}
                             - [ \l_{\db} , W^j ]   \cr
\n_{\a}^{\ \da} \c_{j\da} &=\   [ \c^i_{\a} , W^k ] \e_{ijk}
			      -  [ \l_{\a} , W_j ]  \cr
\n_{\a\da} \n^{\a\da} W_j &=\
- 2 [[ W^i , W_j ], W_i] +  [[ W^i , W_i ], W_j]
- \{ \c_{j\da} , \l^\da\}
 + \half \e_{ijk} \{ \c^i_{\a} , \c^{k\b} \}  \cr
\n_{\a\da} \n^{\a\da} W^j &=\
- 2 [[ W_i , W^j ], W_i]  +  [[ W_i , W^i ], W_j]
- \{ \c^j_{\a} , \l^\a \}
+ \half \e^{ijk} \{ \c_{i\da} , \c_k^\da \}   ,\cr
}\eqno(1)$$
To describe this theory invariantly in the customary superspace with
coordinates
$$\{ x^{\a\da}, \t^{i\a}, \bt_i^{\da}\}, \eqno(2)$$
one introduces gauge-covariant derivatives
$\D_A \equiv \p_A +  A_A = (\n_{\a\db}, \D_{i\a}, \bD^j_{\db}),
i,j = 1,2,3$. The super-connections $A_A$ contain
the above supermultiplet consistently only if the gauge covariant derivatives
are constrained as follows [14, 15]:
$$\eqalign{
   \{\D_{(i\a} ~,~\D_{j)\b}\} =&\ 0  \cr
  \{\bD^{(i}_{\da} ~,~\bD^{j)}_{\db}\} =&\ 0 \cr
  \{\D_{i\a} ~,~\bD^j_{\db}\} =&\  2\delta^j_i \n_{\a\db}  ,\cr}\eqno(3)$$
The crucial message [13] is that these constraints turn out to be
equivalent to the equations of motion (1). Moreover, they take the form
of a Cauchy-Riemann system in an appropriately enlarged base space, as we now
describe.
\vskip 15pt{\bf 3. Harmonic superspace}\vskip 15pt
Let us begin with Euclidean superspace. The coordinates (2) parametrise
the coset of the super Poincare group by its Lorentz subgroup.
In this case the Lorentz group is $SO(4) = SU(2)\times SU(2)$, with
independent $SU(2)$ groups. Factoring, instead, by a {\it subgroup} of the
Lorentz group, yields a correspondingly larger space, a construction which
turns out to be very useful. In particular, factoring by the
$U(1)\times U(1)$ subgroup of the Lorentz group, yields an enlargement of
superspace by a direct product of two 2-spheres, $S^2 = {SU(2)\over U(1)}$.
As coordinates for these spheres we shall use harmonics $u^+_\da , u^-_\da$
and $v^\op_\a , v^\om_\a$ [8], defined up to the respective $U(1)$ phases,
with $(+,-)$ and $(\op,\om)$ being the respective $U(1)$ charges, and
obeying the constraints:
$$ u^{+\da} u^-_\da = 1  ~,~ v^{\op\a} v_\a^{\om} = 1\  ,$$
(or, equivalently, satisfying the completeness relations
$$ u^{+\da} u_\db^{-} - u^{-\da} u_\db^{+} = \d^\da_\db  ~,~
v^{\op\a} v_\b^{\om} - v^{\om\a} v_\b^{\op} = \d^\a_\b \ ).$$
This enlarged space, with spinor harmonics as additional coordinates, is
our \hs.

For the signature (3,1) Minkowski space, the Lorentz group is the simple
group $SL(2,C)$. In this case the analogous construction involves
an enlargement by a coset ${SL(2,C)\over SL(1,C)}$,
so that the charge of the harmonics becomes complex: the $u$ and $v$
harmonics; and the $(+, -)$ and $(\op, \om)$ charges becoming complex
conjugates of each other.
On the other hand, for a signature (2,2) space, the Lorentz group is
$SL(2,R)\times SL(2,R)$, a direct product of two {\it real} groups.
In these noncompact cases there appear richer structures and peculiarities.
We shall not go into subtleties in the present paper. Suffice to say that
in these cases we understand the construction in the sense of a
Wick rotated version of the Euclidean one.

The upshot is that we have additional coordinates and are able to pass
to a basis of \hs with coordinates
$$\{ x^{\pm\op}, x^{\pm\om}, \t^{i\op},\t^{i\om}, \bt_i^{+},
\bt_i^{-}, u^{+}_{\da},u^-_\da, v^{\op}_{\a}, v^{\om}_{\a}\}
\eqno(4)$$ where the
$x$'s and $\t$'s are related to the usual superspace coordinates (2) by
$$  x^{\pm\op} =   x^{\a\da} u^{\pm}_\da  v^{\op}_\a ~,~
    x^{\pm\om} =  x^{\a\da} u^{\pm}_\da  v^{\om}_\a , $$
$$     \t^{i\op} =  \t^{i\a}  v^{\op}_\a ~,~
      \t^{i\om}    =  \t^{i\a}  v^{\om}_\a ~,~
      \bt_i^{\pm} = \bt_i^{\da} u^{\pm}_\da .$$
In virtue of these relations we may recover customary superspace fields
as coefficients in the double harmonic expansions (in both $u$ and $v$ ) of
harmonic superspace fields.
\vskip 15pt{\bf 4. The harmonic superconnections}\vskip 15pt
Now we come to the crucial point. In the coordinates (4)
the constraints (3) are radically simplified: They turn out to be
equivalent to the following set of commutation relations
$$\eqalign{
\{\bD^{+i} ~,~ \bD^{+j} \} =&\ 0\ = \{ \D^{\op }_i ~,~ \D^{\op }_j \}\cr
\{ \bD^{+j} ~,~ \D^{\op}_i \} =&\  2 \n^{+\op},   \cr}\eqno(5)$$
where $\bD^{+i}, \D^{\op}_i $ are gauge-covariant spinorial derivatives and
$\n^{+\op} = \der{x^{-\om}} + A^{+\op} ,$
together with the conditions
$$ \eqalign{[D^{++}, \bD^{+j}] =&\ 0\ =\  [D^{++}, \D^{\op }_i ]  \cr
[D^{++} , \n^{+\op}] =&\ 0\ =  [D^{\op \op} ,  \n^{+\op}] \cr
[D^{\op\op} , \bD^{+j}] =&\ 0\ = [D^{\op\op} , \D^{\op}_i ]\cr}\eqno(6)$$
and the important consistency relation
$$ [ D^{++} , D^{\op \op} ] = 0 , \eqno(7)$$
where $D^{\op\op} ,  D^{++} $ are  harmonic space derivatives which act
on the respective negatively-charged harmonic
space coordinates to yield their positively-charged counterparts, i.e.
$$ \dpp  x^{- \op} =  x^{+ \op} ,\quad \dpp  x^{- \om} =  x^{+ \om} ,\quad
  \dpp u^-_\da = u^+_\da  ,\quad   \dpp  \bt_i^{-} = \bt_i^{+} $$ and
$$ \dopp  x^{\pm\om} =  x^{\pm\op} ,\quad \dopp v^\om_\da = v^\op_\da  ,\quad
   \dopp \t^{i\om} = \t^{i\op}\  ,$$
while giving zero when applied to the respective positively charged
coordinates:
$$ \dpp  x^{+ \op} =  0 ,\quad \dpp  x^{+ \om} =  0 ,\quad
  \dpp u^+_\da = 0  ,\quad   \dpp  \bt_i^+ = 0 $$ and
$$ \dopp  x^{\pm\op} =  0 ,\quad \dopp v^\op_\da = 0  ,\quad
   \dopp \t^{i\op} = 0 \  .$$
Correspondingly, the action on derivatives is given by
$$ [\dpp,\p^{- \op}] = \p^{+ \op} ,\quad [\dpp,\p^{- \om}] = \p^{+ \om} ,
\quad [\dpp, \D^{-i}] = \D^{+i} ,$$
$$ [\dopp,  \p^{\pm\om}] = \p^{\pm\op} ,\quad [\dopp, \D^\om_i] = \D^\op_i,$$
$$ [\dpp,\p^{+ \op}] = [\dpp,\p^{+ \om}] = [\dpp, \D^{+i}] = 0 ,$$
$$ [\dopp,  \p^{\pm\op}] =  [\dopp, \D^\op_i] = 0 .$$
In virtue of these properties of $\dpp,\dopp$ the conditions (6) ensure
that in this basis, the covariant derivatives
$\{ \bD^{+i}, \D^{\op}_i,\n^{+\op} \}$ are homogeneous of degree one in the
correspondingly charged harmonics. The relations (5) mean that these
covariant derivatives take the pure-gauge forms $$\eqalign{
  \D^{+i}     =&\ D^{+i}       -  D^{+i}      \f  \f^{-1}    \cr
  \D^{\op }_i =&\ D^{\op }_i   -   D^{\op }_i  \f  \f^{-1}    \cr
  \n^{+\op}   =&\ \p^{+\op}    -   \p^{+\op}   \f  \f^{-1},    \cr}
\eqno(8)$$
in other words, equations (5) are integrability conditions of the system
$$\eqalign {\D^{+i} \f = 0  \cr
\D^{\op }_i  \f = 0 \cr
 \n^{+\op} \f = 0. \cr} $$
The matrix function $\f$ takes values in the gauge group and is defined up
to the gauge freedom
$$  \f \mapsto e^{-\tau} \f e^{\l} ;\quad
 D^{++}  \tau = 0 =  D^{\op \op} \tau  ,\quad
 D^{+i}  \l = 0 = D^{\op }_i  \l  =  \p^{+\op}  \l ,$$
where $\tau$ and $ \l$ are matrix functions in the gauge algebra.

In contrast to the covariant derivatives (8), the harmonic derivatives
$D^{++}, \dopp$ are `short' (i.e. have no connection)
in this frame. This choice of frame (the `central frame') is actually
inherited from the four-dimensional superspace and is not
the most natural one for harmonic superspace. However, we may pass to
another frame, what we call the `analytic frame', in which  the derivatives
$\{ \D^{+i} , \D^{\op }_i, \n^{+\op} \}$ are  `short' and  $D^{++}, \dopp$
are `long' (i.e. acquire Lie-algebra-valued connections) instead. Namely,
$$\dpp \mapsto \Dpp  =\ \f^{-1}[\dpp] \f = \dpp + \vpp  $$
$$\dopp \mapsto \Dopp  =\ \f^{-1}[\dopp] \f = \dopp + \vopp  ,$$
with the thus acquired harmonic superconnections given by
$$\eqalign{
\vpp =&  \f^{-1}\dpp \f  \cr
\vopp =& \f^{-1}\dopp \f ,\cr}\eqno(9)$$
and the covariant derivatives $\{ \D^{+i} , \D^{\op }_i, \n^{+\op} \}$
lose their connections, i.e. instead of (8) we have, in this analytic frame,
$$\eqalign{
  \D^{+i}     =&\  D^{+i}         \cr
  \D^{\op }_i =&\  D^{\op }_i    \cr
  \n^{+\op}   =&\  \p^{+\op}  .    \cr}$$
\vskip 15pt{\bf 5. The Cauchy-Riemann equivalence}\vskip 15pt
In this analytic frame it is natural to use an analytic basis which
manifestly distinguishes the analytic subspace, analogous to the well known
chiral basis of ordinary superspace which distinguishes the chiral subspace.
Such a basis is defined by the change of coordinates to:
$$ x^{+\om}_A = x^{+\om} +  \bt^+_i\t^{\om i} \  ,\quad
   x^{-\op}_A = x^{-\op} -  \bt^-_i\t^{\op i} $$
$$ x^{-\om}_A = x^{-\om}\   ,\quad  x^{+\op}_A = x^{+\op}\  ,$$
with all other coordinates remaining unchanged. In such an analytic basis the
derivatives occurring the system (5) take the form
$$\eqalign{
D_i^{\op} =& - \der {\t^{\om i}} + \bt_i^{-}\p^{+\op}_A    \cr
\bar D^{+ i} =&\ \der {\bt_i^{-}} - \t^{\om i} \p^{+\op}_A     \cr
\p^{+\op}_A =&\  \der{x^{-\om}_A}                         \cr}\eqno(10)$$
$$ D^{++} = u^{+ \da} \der {u^{- \da}} + x_A^{+ (\op} \p_A^{+ \om)}
+ \bt^+_i \der {\bt_i^{-}} -  \bt^+_i \t^{(\om i}\p_A^{+\op)} $$
$$  D^{\op\op}= v^{\op \a} \der {v^{\om \a}} + x_A^{(+ \op} \p_A^{-) \op}
+\t^{\op i} \der{\t^{\om i}} -  \t^{i\op} \bt^{(+}_i \p_A^{-)\op}\  ,$$
where brackets denote symmetrisations in the corresponding charges.
In the analytic frame the equations (5) become identities for the
derivatives (10) whereas the equations (6) take the form of
generalised Cauchy-Riemann conditions
$$\eqalign{
 {\p \over\p {\bar\t_i^-}}     V^{++} =&\ 0
                            = {\p \over\p {\bar\t_i^-}} V^{\op \op}  \cr
 {\p \over\p {\t^{\om i}}} V^{++} =&\ 0
                          = {\p \over\p {\t^{\om i}}}V^{\op \op} \cr
 {\p \over\p {x_A^{-\om}}}        V^{++} =&\ 0
         = {\p \over\p {x_A^{-\om}}} V^{\op\op}     \cr} \eqno(11)$$
with the consistency relation (7) taking the form of a zero curvature
relation
$$D^{++} V^{\op\op} - D^{\op\op} V^{++} + [V^{++} ,V^{\op\op}] = 0\
.\eqno(12)$$
These equations are merely analytic frame versions of the central frame
equations (5), (6) and (7); and are therefore equivalent to the
constraints (3). The complete dynamical
information of the N=3 Yang-Mills system (1) is therefore coded into
harmonic connections  $\{V^{++} , V^{\op\op}\}$ solving the first order
{\it linear} differential equations (11,12) in \hs. The \sd (resp. \asd)
subsets of solutions simply correspond to v- (resp. u-) independent chiral
(i.e. $\bt$- (resp. $\t$-) independent) solutions of (11). The condition of
v (resp. u) independence being tantamount to the vanishing of one of the
harmonic connections, viz. $\vopp$ (resp. $\vpp$); and chirality implying
the independence of an additional x-variable: $x^{-\op}$ (resp. $x^{+\om}$).
This formulation of the \ssdy equations is described in [9]. The
interlocking of SD and ASD data mentioned at the beginning is clearly
manifest in (11,12).

In order to construct the superconnection satisfying (3), we first need to
recover the `bridge' $\f$ connecting the analytic frame to the central frame
by solving the linear system of equations (9)
$$\eqalign{
 D^{++}    \f  =&\  \f V^{++}           \cr
 D^{\op\op}\f =&\  \f V^{\op\op} \cr}\eqno(13)$$
for arbitrary holomorphic
(i.e. independent of $ \{ x^{-\om} , \t^{\om i} , \bar\t_i^- \}$)
superfields $\{ V^{++} , V^{\op\op} \}$, which enjoy (12) as an
integrability condition. For semi-simple gauge groups it follows from (13)
that the determinant of the bridge obeys the equations
$$ \dpp \det \f = \dopp \det \f = 0. $$
Using a consistent solution $\f$ of this system, we may return to the
central basis in which solutions of the constraints (3) take the form
$$\eqalign{
  A^{+i}       =&\ - D^{+i} \f  \f^{-1}  \cr
  A^{\op }_i=&\ - D^{\op }_i \f  \f^{-1}  \cr
  A^{+\op}   =&\ - \p^{+\op}  \f  \f^{-1} .\cr}$$
These $\{ A^{+i}, A^{+\op} \}$ (resp. $\{ A^{\op }_i , A^{+\op} \}$) are
guaranteed by (5) to be linear in $u$ (resp. $v$) in the central frame,
so the superconnections satisfying (1) afford immediate extraction from
the harmonic expansions
$$\eqalign{
A^{+i}         =&\   u^{+\da}     A^i_\da            \cr
A^{\op }_i  =&\   v^{\op\a} A_{i\a}               \cr
A^{+\op}    =&\   u^{+\da}  v^{\op\a} A_{\a\da}   .\cr} $$
\goodbreak\vskip 15pt{\bf 6. The static limit}\vskip 15pt
The explicit construction is somewhat simpler for the {\it static} case,
which is similar to the three-dimensional $N=6$ Yang-Mills theory considered
in [11].
In these cases the spinor indices $\a$ and $ \da$ as well as two sets of
harmonics get identified and we have the relations to three-dimensional
superspace coordinates:
$$  x^{\pm\pm} =   x^{\a\b} u^{\pm}_\a  u^{\pm}_\b ~,~
    x^{+-} =   x^{\a\b} u^{+}_\a  u^{-}_\b , $$
$$     \t^{i\pm} =  \t^{i\a}  u^{\pm}_\a ~,~
      \bt_i^{\pm} = \bt_i^{\da} u^{\pm}_\da \ ,$$
which supersymmetrise the three dimensional twistor relations of [16].
In this reduced \hs, the harmonic connections $\vpp$ and $\vopp$ become
identified and the somewhat difficult consistency relation (12) disappears,
leaving the simplified CR system
$$\eqalign{
 {\p \over\p {\bt_i^-}}     V^{++} =&\ 0 \cr
 {\p \over\p {\t^{- i}}} V^{++} =&\ 0  \cr
 {\p \over\p {x^{--}}}     V^{++} =&\ 0 ,\cr} $$
in which the imposition of chirality  (i.e. independence of $\bt_i^+$ as
well, which implies $x^{+-}$ independence) corresponds to the Bogomolny
reduction for \sd monopoles (supersymmetrising the construction of [17]).
We shall present explicit solutions elsewhere [5].
\vskip 15pt{\bf 7. Conclusion}\vskip 15pt
We hope to have convinced the reader that the \hs approach is an effective
framework for discussing integrability properties of the $N = 3$ Yang Mills
equations. We find it remarkable that
after sufficient supersymmetrisation non-integrable equations become
integrable and the corresponding quantum theory becomes ultraviolet finite;
and we expect an analogous phenomenon for the higher extended supergravity
theories, which we shall discuss elsewhere.

Acknowledgements. One of us, V.O., would like to acknowledge gratefully
the receipt of a  Humboldt Forschungspreis enabling the performance
of this work at Bonn University and to thank warmly, professors of this
university, V. Rittenberg, G. von Gehlen and E. Sokatchev for discussions
and hospitality.
\goodbreak \vskip 15pt{\bf References }\vskip 15pt
\item{[1]}
L. Sadun, J. Segert, Commun. Math. Phys. 145 (1992) 363;
G. Bor, Commun. Math. Phys. 145 (1992) 393;
H.-Y. Wang, J. Diff. Geom. 34 (1991) 701;
L. Sibner, R. Sibner, K. Uhlenbeck, Proc. Natl. Acad. Sci. USA 86 (1989) 860.
\item{[2]} E. Witten, Phys. Lett. 77B (1978) 394.
\item{[3]} P. Green, J. Isenberg and P. Yasskin, Phys. Lett. 78B (1978) 462.
\item{[4]} M.G. Eastwood, Trans. Amer. Math. Soc. 301 (1987) 615;
 N.P. Buchdahl, Trans. Amer. Math. Soc. 288 (1985) 431;
 J. Harnad, J. Hurtubise, and S. Shnider, Ann. Phys. (N.Y.) 193 (1989) 40.
\item{[5]} C. Devchand and  V. Ogievetsky, to appear.
\item{[6]} R.S. Ward, Phys. Lett. Phys. Lett. 61A (1977) 81.
\item{[7]} Yu. Manin, J. Sov. Math. 2 (1983) 465;
Yu. Manin, {\it Gauge Field Theory and Complex Geometry}, Nauka, Moscow,
1984 (English version, Springer-Verlag, Berlin, 1988).
\item{[8]} M. Evans, F. G\"ursey, V. Ogievetsky, Phys.Rev. D47 (1993) 3496.
\item{[9]} C. Devchand and V. Ogievetsky,  Phys.Lett.B297 (1992) 93;
The matreoshka of supersymmetric self-dual theories, Bonn prepr.
Bonn-HE-93-23, Nucl. Phys. B (in print).
\item{[10]}
A. Galperin, E. Ivanov, S. Kalitzin, V. Ogievetsky, E. Sokatchev,
Class. Quant. Gravity 1 (1984) 469; 2 (1985) 155.
\item{[11]} B. Zupnik, Sov. J. Nucl. Phys. 48 (1988) 744.
\item{[12]} L. Brink, J. Schwarz and J. Scherk, Nucl. Phys. B121 (1977) 77.
\item{[13]} J. Harnad, J. Hurtubise, M. L\'egar\'e and S. Shnider,
Nucl. Phys. B256 (1985) 609; J. Harnad and S. Shnider, Commun. Math.
Phys. 106 (1986) 183.
\item{[14]} R. Grimm, M. Sohnius and J. Wess, Nucl. Phys. B133 (1978) 275.
\item{[15]}  M. Sohnius, Nucl. Phys. B136 (1978) 461.
\item{[16]} R.S. Ward, J. Math. Phys. 30 (1989) 2246.
\item{[17]} O.Ogievetsky, in Group Theoretical Methods in Physics, ed.
H.-D. Doebner et al, Springer Notes in Physics 313 (1988) 548.
\end